\documentclass{article}
\usepackage{hiph-art}
\usepackage{graphicx}
   \newcommand{\vecbm}[1]{\mbox{\boldmath#1}}
\volnumber{19} \issuenumber{1} \edyear{2005}                             
\frompage{000} \topage{000}                                              
\recrevdate{20 October 2005}                                              
\def\rightmark{Microcanonical Thermostatistics}
\def\leftmark{The Proper Treatment of Phase Separation}

\title{Microcanonical Thermostatistics, the basis for a New
Thermodynamics, ``heat can flow from cold to hot'', and nuclear
multifragmentation. The correct treatment of Phase Separation after 150
years of statistical mechanics\footnote{Invited plenary talk at VI Latin
American Symposium on Nuclear Physics and Applications, Iguaz\'u,
Argentina. October 3 to 7, 2005 }}
\authors{D.H.E.Gross
\\[2.812mm]
{\normalsize \hspace*{-8pt}$^1$ Hahn Meitner Institut,\\14109 Berlin,
Germany}}

\abstract{Equilibrium statistics of finite Hamiltonian systems is
fundamentally described by the microcanonical ensemble ($M\!\!E$) [1].
Canonical, or grand-canonical partition functions are deduced from this by
Laplace transform. Only in the  thermodynamic limit are they equivalent to
$M\!\!E$ for homogeneous systems. Therefore $M\!\!E$ is the only ensemble
for non-extensive/inhomogeneous systems like nuclei or stars where the
$\lim_{N\to \infty,\rho=N/V=const}$ does not exist. Conventional canonical
thermo-statistic is inapplicable for non-extensive systems. This has far
reaching fundamental and quite counter-intuitive consequences for
thermo-statistics in general: Phase transitions of first order are signaled
by convexities of $S(E,N,Z,\cdots)$ \cite{gross174}. Here the heat capacity
is {\em negative}.  In these cases heat can flow from cold to hot! The
original task of thermodynamics, the  description of boiling water in heat
engines can be treated. Consequences of this  basic peculiarity for nuclear
statistics as well for the fundamental understanding of Statistical
Mechanics in   general are discussed. Experiments on hot nuclei show all
these  novel phenomena in a rich variety. The close similarity to
inhomogeneous astro physical systems  will be pointed out.}


\makeindex
\begin{document}
\maketitle

\section{Introduction}\label{intro}
Boiling water drives steam engines. Steam engines were the
original motive for proposing Thermodynamics some $170$ years ago.
About $150$ years ago Boltzmann and Gibbs developed Statistical
Mechanics to explain (irreversible) Thermodynamics within
reversible Mechanics. Since then canonical Boltzmann-Gibbs (BG)
statistics is considered to answer this task. However, BG works in
the thermodynamic limit of homogeneous systems. Boiling water is
inhomogeneous. Phase transitions of first order are signaled by
Yang-Lee singularities in BG.  Consequently, BG is inappropriate
for boiling water.
\begin{figure}[h]
\vspace*{10cm}\includegraphics*[bb = 494 0 1 550,angle=-90, width=12 cm,
clip=tru]{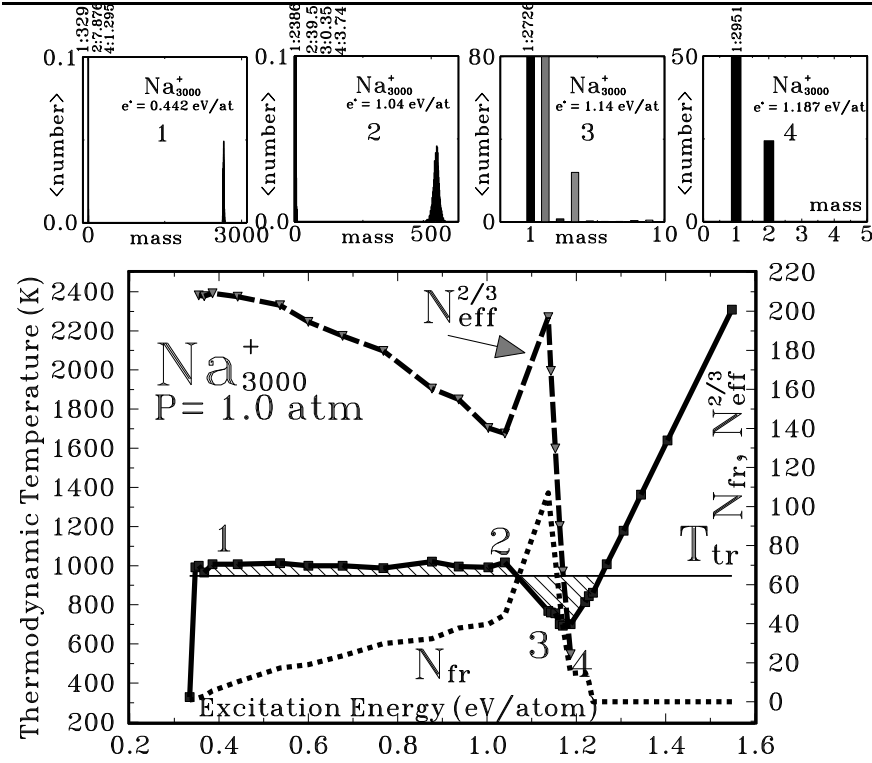}\caption{Microcanonical caloric curve of $3000$ Na
atoms at a pressure of $1$ atm with realistic interaction (c.f.
\cite{gross174}). The temperature $T$ in Kelvin, the energy in eV/atom.
$N_{fr}$ the number of fragments, $N^{2/3}_{eff}=\sum{N_i^{2/3}}$ the
number of surface atoms of dimers and heavier ones. The four inserts give
the fragment distribution at 4 characteristic energies over the intruder.
E.g.: at $0.442$ eV 1:329 means 329 monomers, and 2:7.876 means 7.876
dimers, and 4:1.295 means 1.295 quadrimers on average. Below $e\sim 0.33$
eV $2999$ atoms are condensed in a single liquid droplet and above $e\sim
1.3$ eV nearly all atoms are free ($\sim$ ideal gas). \label{figa}}
\end{figure}

The best way to understand the difference between the canonical and the
microcanonical representation of first order phase transition is perhaps by
studying figure (\ref{figa}), which shows the typical convex intruder of
$S(E)$ between $e\sim 0.35$ and $1.25$ eV/atom. This region gets completely
jumped over by the canonical ensemble. Between $e\sim 0.35$ and $\sim 1.$
eV the temperature (inverse slope of $S(E)$) is approximately constant.
There is a single cluster that evaporates up to $2500$ atoms. This is the
region of the ``compound nucleus for ever'' which is emphasized so much by
Moretto \cite{bowman87}. However, its temperature is {\em higher} than the
canonical transition temperature $T_{tr}$ and these events are inaccessible
to the canonical ensemble.

At $\sim 1.1$ eV the single cluster explodes suddenly into on average
$2726$ monomers, $80$ dimers, about $20$ quadrimers and a few heavier
fragments. This corresponds to nuclear multi-fragmentation. However, as the
total charge of the system is only $1^+$ there is no Coulomb force, and
fission is missing. One can also clearly see that the total surface of the
fragments $\propto N^{2/3}_{eff}$ rises steep in the fragmentation region.
This leads to the steep drop of the temperature (backbending).

\section{Short way to entropy}\label{entropy}
Entropy, S, is the fundamental entity of thermodynamics which
distinguishes thermodynamics from all other physics. Therefore,
its proper understanding is essential. The understanding of
entropy is sometimes obscured by frequent use of the
Boltzmann-Gibbs canonical ensemble, and the thermodynamic limit.
Also its relationship to the second law is often beset with
confusion between external transfers of entropy $dS_e$ and its
internal production $dS_i$ \cite{gross216}.

Macroscopic systems are controlled by a few macroscopic control
parameters $M$ like energy $E$, particle number $N$, and volume
$V$. All $6N-M$ degrees of freedom remain unknown. Consequently,
Thermodynamics describes all systems with the same macroscopic
parameters $M$ simultaneously. It can only give probabilistic
information about the typical behaviour of all systems in the
microcanonical ensemble.

Boltzmann defined the entropy of an isolated system
\begin{equation}
\fbox{\fbox{\vecbm{S=k*lnW}}}
\end{equation}
as written on Boltzmann's tomb-stone. I.e. $S$ measures the
size of the (microcanonical) ensemble.

This has a very simple interpretation: The size of the ensemble i.e. the
number $W$ of cells of size $(2\pi\hbar)^{3N}$ of different initial values
of positions $q_i$ and momenta $p_i$ consistent with the $M$ control
parameter is measured by the entropy $S$. If we would know all $6N$
positions and momenta, $W$ would be one and the entropy $S=0$.

\section{No phase separation without convex non-extensive {\boldmath$S(E)$}}\label{convex}
The weight $e^{S(E)-E/T}$ of configurations with energy E in the
definition of the canonical partition sum
\begin{equation}
Z(T)=\int_0^\infty{e^{S(E)-E/T}dE}\label{canonicweight}
\end{equation} becomes here {\em bimodal}, at the transition temperature it has
two peaks, the liquid and the gas configurations which are separated in
energy by the latent heat. Consequently $S(E)$ must be convex (like
$y=x^2$) and the weight in (\ref{canonicweight}) has a minimum between the
two pure phases. Of course, the minimum can only be seen in the
microcanonical ensemble where the energy is controlled and its fluctuations
forbidden. Otherwise, the system would fluctuate between the two pure
phases by an, for macroscopic systems even macroscopic, energy $\Delta
E\sim E_{lat}\propto N$ of the order of the latent heat. I.e. {\em the
convexity of $S(E)$ is the generic signal of a phase transition of first
order} and of phase-separation\cite{gross174}.  Such macroscopic energy
fluctuations and the resulting negative heat capacity are already early
discussed in high-energy physics by Carlitz \cite{carlitz72}.

\section{Application in astrophysics}
The necessity of using ``extensive'' instead of ``intensive'' control
parameter is explicit in astrophysical problems. E.g.: for the description
of rotating stars one conventionally works at a given temperature  and
fixed angular velocity $\Omega$ c.f. \cite{chavanis03}. Of course in
reality there is neither a heat bath nor a rotating disk. Moreover, the
latter scenario is fundamentally wrong as at the periphery of the disk the
rotational velocity may even become larger than velocity of light.
Non-extensive systems like astro-physical ones do not allow a
``field-theoretical'' description controlled by intensive fields !

E.g. configurations with a maximum of random energy on a rotating
disk, i.e. at fixed rotational velocity $\Omega$:
\begin{equation}
E_{random}=E-\frac{\Theta\Omega^2}{2} -E_{pot}
\end{equation} and consequently with the largest entropy are the ones
with smallest moment of inertia $\Theta$, compact single stars. Just the
opposite happens when the angular-momentum $L$ and not the angular velocity
$\Omega$ are fixed:\begin{equation} E_{random}=E-\frac{L^2}{2 \Theta}
-E_{pot}.
\end{equation}Then configurations with large moment of inertia are
maximizing the phase space and the entropy. I.e. eventually double or multi
stars are produced, as observed in reality.

\begin{figure}[h]
\includegraphics[bb =0 0 511 353,width=10cm,angle=0,clip=true]{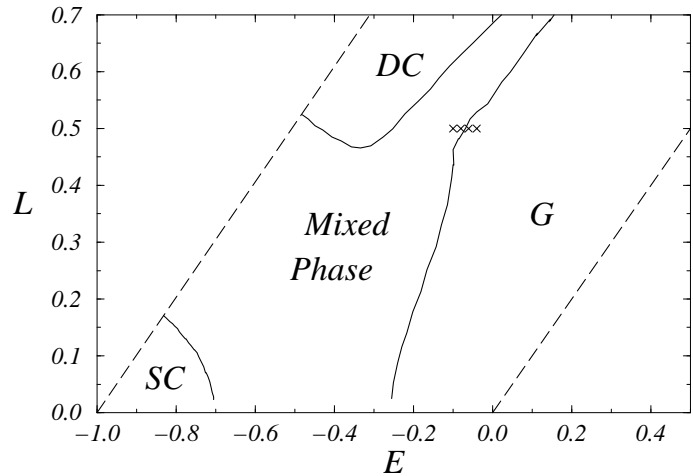}
 \caption{Phase diagram of rotating self-gravitating systems
in the energy-angular-momentum $(E,L)$-plane. DC: region of
double-stars, G: gas phase, SC: single stars. In the mixed region
one finds various exotic configurations like ring-systems in
coexistence with gas, double stars or single stars. In this region
of phase-separation the heat capacity is negative and the entropy
is convex. The dashed lines $E-L=-1$ (left) and $E=L$ (right)
delimit the region where systematic calculations were carried out.
At a few points outside of this strip some calculations like the
left of fig.(\ref{cover}) were also done. \label{phased}}
\end{figure}

In figure \ref{phased} one clearly sees the rich and realistic
microcanonical phase-diagram of a rotating gravitating system controlled by
the ``extensive'' parameters energy and angular-momentum. \cite{gross187}

\begin{figure}[h]
\includegraphics*[bb = 93 401 519 592,
angle=-0, width=10cm, clip=tru]{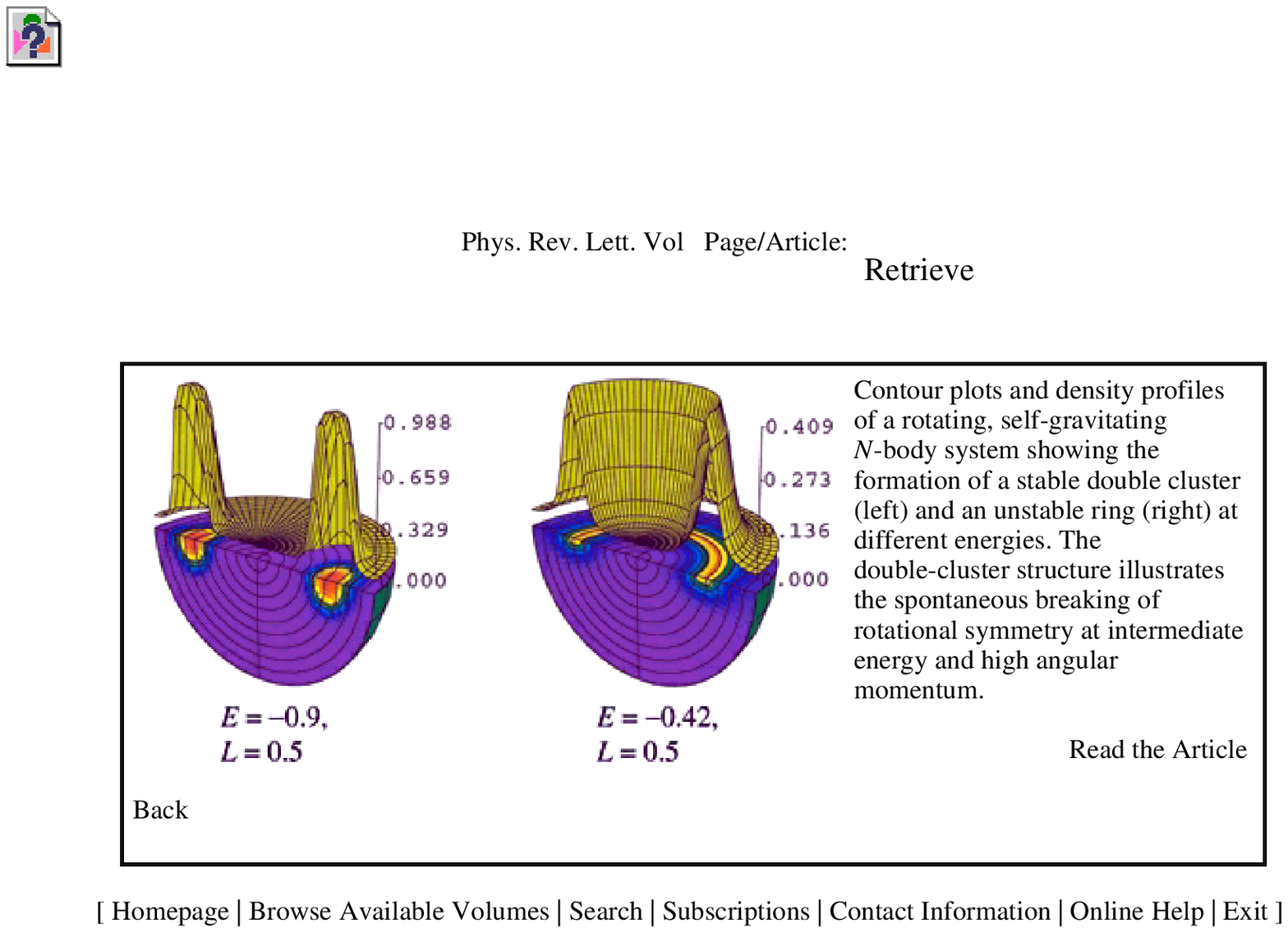}
\caption{Rotating multi-star-systems. The left shows a rotating double-star
system in the DC phase. This is an inhomogeneous phase, analog to nuclear
multifragmentation. The right is one of the many different, unstable,
configurations existing in the mixed phase with negative heat capacity.
Here the system fluctuates between such ring systems, systems of stars
rotating around a central star but also mono-stars and eventual gas. This
region is very interesting but must still be more investigated. Cover page
of Phys.Rev.Lett. vol 89, (July 2002)\label{cover}}
\end{figure}

\begin{thebibliography}{1}

\bibitem{gross174}
D.H.E. Gross.
\newblock {\em Microcanonical thermodynamics: Phase transitions in ``Small''
  systems}, volume~66 of {\em Lecture Notes in Physics}.
\newblock World Scientific, Singapore, 2001.

\bibitem{bowman87}
D.R. Bowman, W.J. Kehoe, R.J. Charity, M.A. McMahan, A.~Moroni, A.~Bracco,
  S.~Bradley, I.~Iori, R.J. McDonald, A.C. Mignerey, L.G. Moretto, M.N.
  Namboodiri, and G.J. Wozniak.
\newblock Complex fragment emission at 50 Mev/u - compound nuclei forever.
\newblock {\em Phys. Lett.}, B 189:282, 1987.

\bibitem{gross216}
D.H.E. Gross.
\newblock On the foundation of thermodynamics by microcanonical
  thermostatistics. The microscopic origin of condensation and phase
  separations.
\newblock  http://arXiv.org/abs/cond--mat/0509202, (2005).

\bibitem{carlitz72}
R.D. Carlitz.
\newblock Hadronic matter at high density.
\newblock {\em Phys.Rev.D}, 5:3231--3242, 1972.

\bibitem{chavanis03}
P.H. Chavanis and M.~Rieutord.
\newblock Statistical mechanics and phase diagrams of rotating self-gravitating
  fermions.
\newblock {\em Astron. Astrophys.}, 412:1, arXiv:astro--ph/0302594, 2003.

\bibitem{gross187}
E.V. Votyakov, H.I. Hidmi, A.~De Martino, and D.H.E. Gross.
\newblock Microcanonical mean-field thermodynamics of self-gravitating and
  rotating systems.
\newblock {\em Phys.Rev.Lett.}, 89:031101--1--4;
  http://arXiv.org/abs/cond--mat/0202140, (2002).

\end{thebibliography}

\end{document}